# Revealing the molecular structures of $\alpha$-Al$_2$O$_3$(0001) –water interface by machine learning based computational vibrational spectroscopy

Xianglong Du[1], Weizhi Shao[2,3], Chenglong Bao[2], Linfeng Zhang[4,3], Jun Cheng[1,6,7*], Fujie Tang[5,6*]

1. State Key Laboratory of Physical Chemistry of Solid Surfaces, iChEM, College of Chemistry and Chemical Engineering, Discipline of Intelligent Instrument and Equipment, Xiamen University, Xiamen 361005, China
2. Yau Mathematical Sciences Center, Tsinghua University, Beijing 100084, China
3. AI for Science Institute, Beijing 100080, China
4. DP Technology, Beijing 100080, China
5. Pen-Tung Sah Institute of Micro-Nano Science and Technology, Xiamen University, Xiamen 361005, China.
6. Laboratory of AI for Electrochemistry (AI4EC), Tan Kah Kee Innovation Laboratory (IKKEM), Xiamen 361005, China
7. Institute of Artificial Intelligence, Xiamen University, Xiamen 361005, China

*Author to whom correspondence should be addressed: chengjun@xmu.edu.cn; tangfujie@xmu.edu.cn

## Abstract

Solid-water interfaces are crucial to many physical and chemical processes and are extensively studied using surface-specific sum-frequency generation (SFG) spectroscopy. To establish clear correlations between specific spectral signatures and distinct interfacial water structures, theoretical calculations using molecular dynamics (MD) simulations are required. These MD simulations typically need relatively long trajectories (a few nanoseconds) to achieve reliable SFG response function calculations via the dipole-polarizability time correlation function. However, the requirement for long trajectories limits the use of computationally expensive techniques such as *ab initio* MD (AIMD) simulations, particularly for complex solid-water interfaces. In this work, we present a pathway for calculating vibrational spectra (IR, Raman, SFG) of solid–water interfaces using machine learning (ML)-accelerated methods. We employ both the dipole moment-polarizability correlation function and the surface-specific velocity-velocity correlation function approaches to calculate SFG spectra. Our results demonstrate the successful acceleration of AIMD simulations and the calculation of SFG spectra using ML methods. This advancement provides an opportunity to calculate SFG spectra for the complicated solid-water systems more rapidly and at a lower computational cost with the aid of ML.

1. Introduction

Oxide-water interfaces are ubiquitous on Earth and play pivotal roles in numerous fields including catalysis, electrochemistry, and geochemistry.[1-3] At these interfaces, water molecules can undergo dissociation or physical absorption, underscoring their significance.[4-6] A molecular-level comprehension of the hydrogen bond (H-bond) network among water molecules near oxide surfaces is essential for understanding these processes. Experimental methods such as X-ray reflectivity and neutron scattering provide valuable structural insights into oxide surface terminations and the initial water layers at various oxide-water interfaces.[7, 8] However, due to the low mass and charge density of hydrogen atoms, these techniques are inadequate for directly probing the H-bond network of interfacial water. Vibrational spectroscopy, such as infrared (IR) and Raman spectroscopy, offers a complementary approach by probing the OH vibrational mode, which correlates with H-bond strength.[9-11] Nevertheless, these techniques face challenges in directly studying interfacial water due to signal interference from bulk water.[12]

To this end, sum-frequency generation (SFG) spectroscopy is uniquely suited due to two primary advantages: molecular specificity (particularly sensitivity to H-bonding strength and molecular orientation) and interface specificity.[13-16] In an SFG experiment, IR and visible laser fields are combined to generate the sum frequency of these two fields. The signal is enhanced when the IR frequency resonates with the molecular vibration, providing specificity to the molecular structure.[13] Furthermore, Heterodyne-Detected SFG spectroscopy yields complex-valued $\chi^{(2)}$ spectra, where the sign of the imaginary part of the $\chi^{(2)}$ spectrum (Im $\chi^{(2)}$) indicates the absolute orientation of water molecules (*up- or down-orientation*). SFG signals are non-zero only when centro-symmetry is broken, such as at the interfaces.[14, 15] Signals from the bulk water are naturally excluded due to the SFG selection rule.[16] Consequently, the SFG technique is ideally suited for probing molecular structures at various interfaces, including water-air[17-19] and oxide-water interfaces[20-22].

Connecting the experimental SFG signals to the molecular structures of interfacial water is challenging due to the complex nature of interfaces.[23] Molecular dynamics (MD) simulations serve as a powerful tool to investigate the microscopic structure of water molecules in contact with solids.[24-28] By combining MD simulations with SFG calculations, one can analyze experimental details and obtain clear structural information about interfacial water. Given the complex interactions of atoms at solid-water interfaces, classical force field models, which depend heavily on the functions used in force field modeling and fitting procedures, may be insufficient.[16] *Ab initio* molecular dynamics (AIMD) simulations, on the other hand, provide detailed structural information at the atomic scale since the molecular forces are calculated using electronic structure theory.[29-31] Despite AIMD's advantage of being parameter-free when modeling solid-water interfaces, it faces significant challenges due to its high computational cost. A few nanoseconds of MD trajectory would be required to calculate the SFG spectra using the dipole moment-polarizability correlation function ($\mu$-$\alpha$) approach.[32] This method necessitates obtaining the dipole moment and polarizability on-the-fly during AIMD simulations.[33] To overcome these limitations, the surface-specific velocity-velocity correlation function (ssVVCF) approach is developed.[34] This method predicts SFG spectra based on the velocity of AIMD trajectories and empirical parameter approximations. Consequently, only a few hundred picoseconds of trajectory are required, and direct calculation of dipole moment and polarizability is unnecessary. Nevertheless, this method can only study the vibrational modes of specific bonds. Previous research has primarily focused on the vibrational stretching modes of O-H (O-D) bonds,[35, 36] making it difficult to study the vibrational modes of other types of bonds at various complicated interfaces. Thus, both methods have drawbacks in predicting SFG spectra.

Fortunately, the development of machine learning (ML) has made it possible to overcome the shortcomings of these methods. ML potential models can fit potential energy surfaces with first-principle accuracy using a small amount of training data containing potential energy and force,[37-39] effectively expanding the size and time range of AIMD simulations.[40, 41] Thus, combining ML potential models with the ssVVCF approach allows for the prediction of SFG spectra in more complex systems, such as the air-salt solution interface[42] and graphene-water interface.[43] Furthermore, ML models can predict dipole moments and polarizabilities,[44-46] which are valuable for predicting the IR spectra of confined water.[47] Notably, there are some recent works which aim to employ the ML models to predict the dipole moments and polarizabilities with the equivariant dielectric response (MACE) model[46, 48], SA-GPR model[49] as well as the Deep Wannier method[50] to calculate the SFG spectra for water/air or ice/water interfaces. Here, we train several ML models to calculate the SFG spectra of the complicated solid-water interfaces with the rigorous dipole moment-polarizability correlation function ($\mu$-$\alpha$) approach. We will systematically study the molecular structures of the $\alpha$-$Al_2O_3$(0001)–water interface with the vibrational spectroscopy (IR, Raman, and SFG) calculations using ML models to demonstrate the ability of the our methods for the complicated solid-water interfaces. We take the $\alpha$-$Al_2O_3$(0001) surface as a prototype since it is the most stable crystalline surface among the different phases of aluminum oxide.[51] Furthermore, there are many studies calculating the SFG spectra of $\alpha$-$Al_2O_3$(0001)–water interface based on AIMD,[52-54] it can benchmark the results of our methods.

The organization of the rest of the paper is as follows: In Sect. 2, we present the calculation details, including how to obtain the MD trajectory and how to calculate the vibrational spectra. Sect. 3 shows the results obtained based on the MLMD trajectory, including the molecular structures and different vibrational spectra. Finally, the conclusion is given in Sect. 4.

## 2. Computational methods
### 2.1 Molecular dynamic simulation details
#### 2.1.1 Interface model

The cell dimensions of $\alpha$-$Al_2O_3$(0001)–water interface model are 9.514 × 8.239 × 43.1389 Å. The (0001) interface contains six layers of aluminum atoms and seven layers of oxygen atoms, and surface oxygen atoms are all saturated with hydrogen, so each face has 12 OH groups. We placed 68 water molecules in the gap for the (0001) interface, to make sure the density of bulk phase water is close to 1 g/cm$^3$. A snapshot of $\alpha$-$Al_2O_3$(0001)–water slab is shown in Fig. 1.

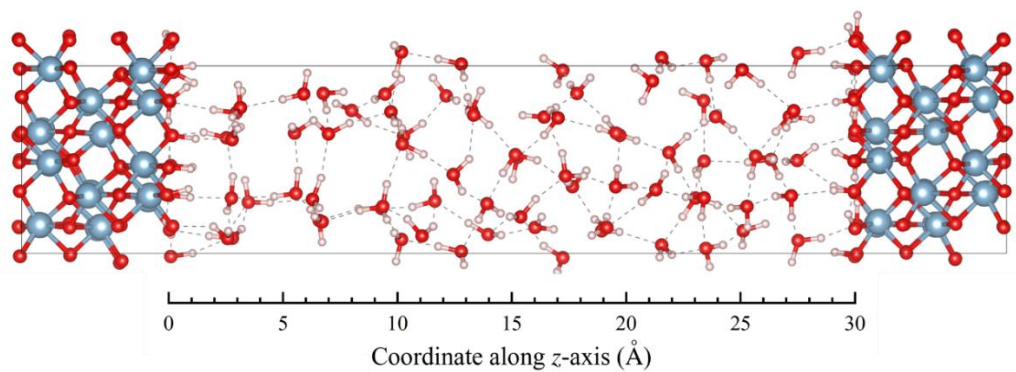

Figure 1. The simulation cell used for $\alpha$-$Al_2O_3$(0001)–water interface model. The z-axis is parallel to the surface normal, and its zero point corresponds to the average position of surface oxygen atoms of the one surface. The distance between the upper surface and lower surface is approximately 30 Å.

#### 2.1.2 DFT settings

All the DFT calculations were performed with the mixed Gaussian and plane-wave basis set using CP2K/QUICKSTEP package.[55] The Perdew-Burke-Ernzerhof (PBE) functional was used to describe the exchange-correlation energies, and the dispersion correction was applied in all calculations with the Grimme D3 method.[56, 57] The

Gaussian basis set was set to double-ζ with one set of polarization functions (DZVP-MOLOPT-SR-GTH) , and the energy cutoff for wavefunction was set to 400 Ry.[58, 59] The core electrons were described by Goedecker-Teter-Hutter (GTH) pseudopotentials.[60] Orbital transformation (OT) algorithm was used to optimize SCF wavefunction,[61] while SCF convergence threshold was set to $1\times10^{-6}$ a.u.. We used the grid interpolation scheme for electron density using the keyword option XC_DERIV NN10_SMOOTH and XC_SMOOTH_RHO NN10.

### 2.1.3 ML models training

We trained various ML models to accelerate the calculation of vibrational spectra using the DeePMD-kit package.[62, 63] First, we trained the Deep Potential (DP) model to accelerate the simulation of the molecular structures,[64] which predicts the potential energy and force of structures. The training data set for the DP model was prepared using a workflow implemented in the DP-GEN package.[65] We employed the similar training approach used in Ref. 41. Here, we briefly discussed the details of the training. We firstly randomly sample 50 structures from the trajectory to train four initial models. Then, three consecutive steps, namely, *exploration*, *labeling*, and *training*, are repeated until MLP prediction reaches sufficient accuracy. The exploration stage of DP-GEN was at the thermodynamic conditions of 330 K, 430 K and 530 K NVT ensemble. The upper and lower trust bounds for model deviation are set as 0.40 and 0.20 eV/Å, respectively. After the model is converged, the systems that included during the exploration stage of DP-GEN were the α‑$Al_2O_3$ (0001)–water interface of 821 structures.

After obtained the molecular structures, we trained two Deep Wannier (DW) models,[44, 45] one for the prediction of the dipole moment and another for the prediction of the polarizability. The structures used to train the DW models were the same as those obtained by iterating in the DP-GEN package. To train the DW models, we summed the four neighboring Wannier centers that belong to the nearest oxygen atom separately. This allowed us to obtain the Wannier centroids (WCs). We then calculated the positions of each WC relative to the nearest oxygen atom. These relative positions correspond to the training data set of the dipole moment. The changes of the WCs under the external electric fields correspond to the polarizability used for training.

**2.1.4 MLMD setting**

After obtained the DP models, we ran MLMD simulations to sample the molecular configurations. All MD simulations were performed in the *NVT* ensemble with the LAMMPS package.[66] We had run MD simulations at three different temperatures (330 K, 360 K and 390 K) with Nose-Hoover chain thermostat and the number of chains is 3.[67, 68] The temperature damping parameter was set to 1 ps, and a 0.5 fs time step was used to integrate the equation of motion. To ensure a better sampling of the molecular configurations, we performed separate 500 ps simulations based on five different initial coordinates for the simulations. Additionally, to improve the convergence of the SFG spectra, we ran five additional 500 ps trajectories at 330 K, considering that most of the SFG spectra were calculated at this temperature.

**2.1.5 ML models validation**

In this section, we will validate the accuracy of the ML models used in this work. The ML models were validated using 100 structures from the trajectories that were not included in the training data set. We computed the energy, forces, Wannier center and related polarizability with DFT and NN models, as shown in Fig. 2. As one can see, Fig. 2 (a) and (b) show that the DP model accurately predicted both energy and force. The energy and force error were evaluated using root mean square error (RMSE). The energy RMSE is $1.55 \times 10^{-1}$ eV, and the force RMSE is $6.90 \times 10^{-2}$ eV/Å. These meeting the prediction accuracy standard of DP model, which requires the force RMSE to be less than $1.00 \times 10^{-1}$ eV/Å. To validate the accuracy of the DW model in predicting dipole moment, we considered the prediction accuracy of the relative positions of WC (ΔWC) instead of the dipole moment. Fig. 2 (c) shows that the DW model successfully predicts the ΔWC, indicating that the dipole moment of trajectories can be accurately predicted. For the prediction of polarizability, it is undesirable as shown in Fig. 2 (d), unlike the DW model reported in the Ref. 45 which has better prediction for the isotropic part of polarizability rather than the anisotropic part. Note that this difference in accuracy may be due to various factors such as the structures, DFT calculational software, functional, etc. But later we will demonstrate that such a prediction accuracy yields reliable Raman and SFG spectra. Moreover, we compared the structural properties predicted by AIMD and DPMD, as illustrated in Fig. 3. The negligible differences between the structural properties obtained from AIMD and DPMD further validate the accuracy of our DPMD model.

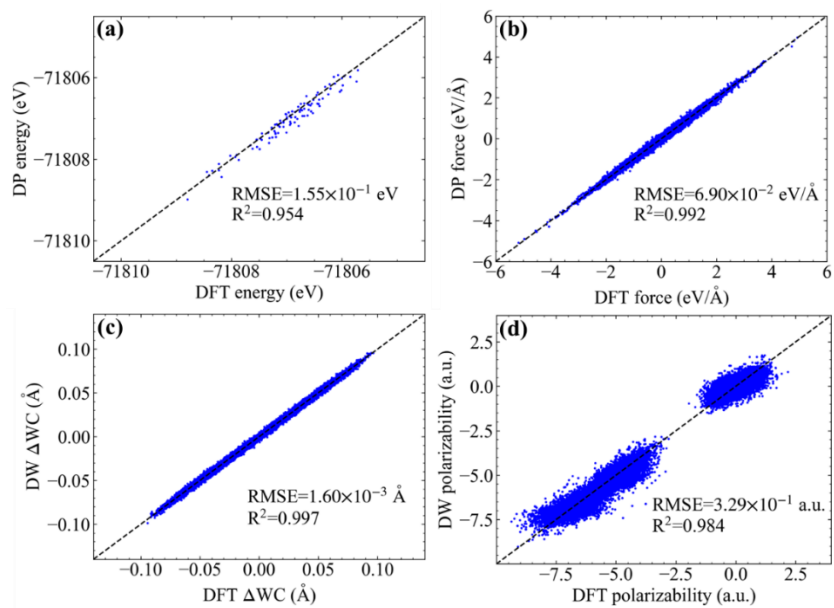

Figure 2. The results of ML models testing in (a) energy, (b) force, (c) dipole moment and (d) polarizability.

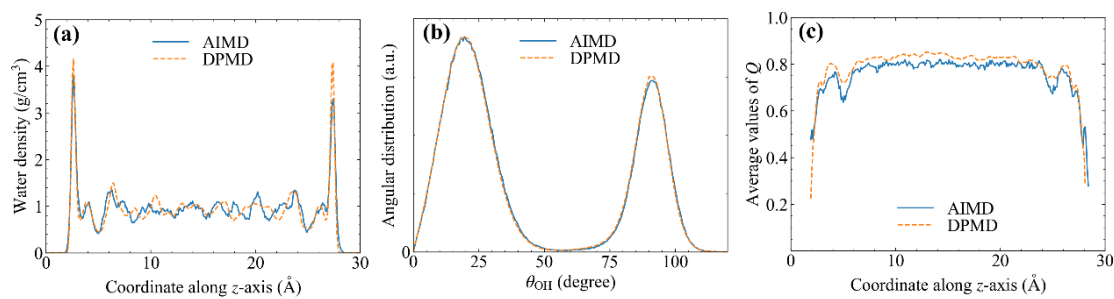

Figure 3. (a) Density profile of water along the $z$-axis of AIMD and DPMD. (b) Angular distribution of surface OH groups of AIMD and DPMD, the angle is defined between the OH group and surface normal. (c) Distribution of $Q$ of water molecules in different layers of AIMD and DPMD, tetrahedral order parameter ($Q$) is defined in Eq. 16.

## 2.2 Spectral calculation methods

### 2.2.1 IR spectra

The formula used to calculate IR spectra of water molecules is shown below:[69]

$$\alpha(\omega)n(\omega) = \frac{2\pi\beta}{3cV} \int_{t=0}^{\infty} e^{-i\omega t} dt \sum_{i=1}^{N} \sum_{j=1}^{N} \langle \dot{\mu}_i(0)\dot{\mu}_j(t) \rangle \tag{1}$$

here we use the product of the absorption cross section $\alpha(\omega)$ and the refractive index $n(\omega)$ instead of intensity. Where $\omega$, $c$, $\beta=1/k_BT$, $\dot{\mu}$ and $V$ are the frequency, speed of light, inverse temperature, the change rate of dipole moment against time, and volume of the system, respectively. $i$, $j$ represent the index of water molecules, and $N$ is the total number of molecules. When we split the dipole moment correlation function into intramolecular and intermolecular contributions, the two parts are expressed as follows:

$$C_{\text{intra}}(t) = \sum_{i=1}^{N} \langle \dot{\mu}_i(0)\dot{\mu}_i(t) \rangle \tag{2}$$

$$C_{\text{inter}}(t) = \sum_{i=1}^{N} \sum_{j\neq i}^{N} S_r(r_{ij}) \langle \dot{\mu}_i(0)\dot{\mu}_j(t) \rangle \tag{3}$$

$$S_r(r_{ij}) = \begin{cases} 0, r_{ij} > r_c \\ 1, r_{ij} \leq r_c \end{cases} \tag{4}$$

The distance switching function $S_r(r_{ij})$ is used to consider the different cross-correlation cutoff, where $r_{ij}$ is the distance between the $i$th and $j$th oxygen atoms, and $r_c$ is the cross-correlation cutoff. For the intermolecular contributions, we only select the molecules within 6 Å of one considered molecule (based on the distance between oxygen atoms). This cutoff value has been shown to effectively consider intermolecular contributions in the IR spectra of water.[10] To calculate the dipole moment, we employed the maximally localized Wannier functions (MLWFs) scheme.[70] This scheme has been shown to be effective in calculating the IR spectra of condensed phases.[69] In this scheme, the Wannier centers represent the positions of bonds and lone pairs of the molecule, which is corresponding to the negative charge centers, while the atomic nucleus corresponds to the positive charge centers. Considering that we used the WC to represent the dipole moment, the dipole moment expressed in this way is shown below:

$$\mu_{H_2O,i} = e\left(6r_O^i + r_{H,1}^i + r_{H,2}^i - 2\sum_{j=1}^{4} r_{W,j}\right) \tag{5}$$

$$\mu_{\text{OH},i} = e\left(6r_O^i + r_H^i + 0.5r_{\text{Al},1} + 0.5r_{\text{Al},2} - 2\sum_{j=1}^{4} r_{\text{W},j}\right) \qquad (6)$$

where $e$ is the elemental charge, $r_H$, $r_O$ and $r_{Al}$ are the positions of atomic nucleus, $r_W$ is the Wannier center, $i$ and $j$ are the index of molecule and Wannier center. So that we can get the dipole moment of each water molecule and surface OH group.

### 2.2.2 Raman spectra

For Raman spectra simulation, we just considered the bulk phase water. The formulas for Raman spectra simulation are shown below:[11]

$$R_{\text{iso}}(\omega) \propto \frac{\hbar\omega}{kT}\int_{t=0}^{\infty} e^{-i\omega t}dt \sum_{i=1}^{N}\sum_{j=1}^{N}\langle \overline{\alpha_i}(0)\overline{\alpha_j}(t)\rangle \qquad (7)$$

$$R_{\text{aniso}}(\omega) \propto \frac{\hbar\omega}{kT}\int_{t=0}^{\infty} e^{-i\omega t}dt \sum_{i=1}^{N}\sum_{j=1}^{N}\left\langle \frac{2}{15}\text{Tr}\beta_i(0)\beta_j(t)\right\rangle \qquad (8)$$

here we decompose the polarizability tensor into an isotropic part and an anisotropic part for the fluid system. $\bar{\alpha}$ and $\beta$ are the isotropic and anisotropic components of the polarizability tensor $\alpha$: $\bar{\alpha}$ = (1/3)Tr$\alpha$ and $\beta$ = $\bar{\alpha}$ − $\alpha$ I, where I is the identity tensor, and Tr denotes a trace operator. $i$, $j$ represent the index of water molecules and $N$ is the total number of considered water molecules. To split the isotropic polarizability correlation function into intramolecular and intermolecular contributions, we followed the same method as described for the dipole moment correlation function.

There are two most common approaches used to calculate the polarizability. One is density functional perturbation theory (DFPT),[71] which solves the self-consistent response equations for the electrons to obtain polarizability, the other is to calculate the changes of dipole moment under the small but finite electric fields.[72, 73] In this case, the latter approach was chosen based on the MLWFs scheme. The electric fields were applied separately along the +x, +y, and +z axes in practice. The changes in the dipole moment were then calculated with and without the applied electric fields in each direction. The polarizability tensor of each molecule was determined by the variation of the three dipole moment vectors.

### 2.2.3 SFG spectra

We had calculated the SFG spectra based on the $\mu$-$\alpha$ approach and the ssVVCF approach, respectively. [34, 74]

For the μ-α approach, the formulas we used to obtain the resonant part of the second-order susceptibility $\chi^{(2)}$ is as follows:[32, 75]

$$\chi_{ijk}^{(2),\text{Res}}(\omega) = \frac{i\omega}{k_bT}\int_{t=0}^{\infty}e^{-i\omega t}dt$$

$$\sum_n S_z^3(z_n(0))\alpha_{ij,n}(t)\cdot\mu_{k,n}(0) + \sum_n\sum_{m\neq n}S_z^2(z_n(0))S_z(z_m(0))S_r(r_{mn}(0))\alpha_{ij,n}(t)\cdot\mu_{k,m}(0) \quad (9)$$

here the indices *i, j, k* represent the Cartesian components *x, y,* and *z*, and index *m, n* to water molecules. *μ* and *α* are the dipole moment and polarizability for each water molecule and surface OH group. The first term represents the intramolecular contribution, while the second term represents the intermolecular contribution. The switching function $S_z(z_n)$ is employed to avoid the cancellation of two opposite interfaces. Here, $z_c$ corresponds to the thickness of bulk phase water and transition region, and $z_n$ corresponds to the *z*-coordinate of the *n*th oxygen atom. $S_r(r_{mn})$ is the same as $S_r(r_{ij})$ in the dipole moment correlation function.

$$S_z(z_n) = \text{sign}(z_n) \times \begin{cases} 0, & |z_n| \leq z_c \\ 1, & |z_n| > z_c \end{cases} \quad (10)$$

For the ssVVCF approach, the SFG response can be written as:[34]

$$\chi_{aac}^{(2),\text{Res}}(\omega) = \frac{Q(\omega)\mu'(\omega)\alpha'(\omega)}{i\omega^2}\int_{t=0}^{\infty}e^{-i\omega t}dt\left\langle\sum_{i,j}S_r(r_{ij}(0))g_z(z_i)\dot{r}_{z,i}^{OH}(0)\frac{\vec{r}_i^{OH}(t)\cdot\vec{r}_j^{OH}(t)}{|\vec{r}_i^{OH}(t)|}\right\rangle$$

(11)

here index *i, j* represent the index of OH chromophore, $S_r(r_{ij})$ is the same as mentioned above, and $g_z(z_i)$ is the truncation function used to select the interfacial water:

$$g_z(z_i) = \begin{cases} 0, & z_i \geq z_d \\ 1, & z_i < z_d \end{cases} \quad (12)$$

where $z_d$ is the *z*-coordinate of the selected dividing surfaces and $z_i$ is the *z*-coordinate of the *i*th oxygen atom. The transition dipole moment and polarizability were considered by using the frequency dependent transition dipole moment (*μ'(ω)*) and polarizability (*α'(ω)*):[76, 77]

$$\mu'(\omega) \equiv \left(1.377 + \frac{53.03(3737-\omega)}{6932.2}\right)\mu^0 \quad (13)$$

$$\alpha'(\omega) \equiv \left(1.271 + \frac{6.287(3737-\omega)}{6932.2}\right)\alpha^0 \quad (14)$$

where *ω* is in cm$^{-1}$. $Q(\omega)$ is the quantum correction factor given by:[78]

$$Q(\omega) = \frac{\beta\hbar\omega}{1-exp(-\beta\hbar\omega)} \tag{15}$$

where $\beta=1/k_BT$ is the inverse temperature.

## 3. Results and discussion

### 3.1 Structural property

In this section, we will discuss the molecular structure of $\alpha$-$Al_2O_3$(0001)–water interface obtained from MLMD trajectories.

#### 3.1.1 Density profile

First, we calculated the density profile of water at different temperatures, as shown in Fig. 3 (a). It is observed that the average distance from the water molecules to the surface is not less than 2.0 Å. The maximum water density is found to be at about 2.6 Å from the interface, and increasing the temperature results in a decrease in the maximum value without affecting the positions corresponding to the maximum value. The density profile reveals two prominent peaks at approximately 4.0 Å and 6.5 Å from the interface, which are slightly affected by temperature. Note that the PBE-D3 method is known to overestimate the H-bond strength, therefore, it will lead to a slightly overstructured water at lower temperature.[79] Thus, we classify the water molecules corresponding to these peaks as transition regions. Based on the density profile, we divide the interfacial water into different water layers: The first, second- and third-layers of water molecules correspond to the range of 2 Å < z < 3 Å, 3 Å < z < 5 Å and 5 Å < z < 8 Å from the interface, respectively. As for the bulk phase water, we toke the water molecules in a slab with the thickness of 6 Å near the middle range.

#### 3.1.2 Angular distribution

The distribution of surface OH tilt angles at different temperatures was calculated, as shown in Fig. 3 (b). The peak near 90° corresponds to OH groups parallel to the surface, while the peak at 20° corresponds to OH groups nearly perpendicular to the surface. The former is able to form two H-bonds with the surrounding surface OH groups as a H-bond donor, whereas the latter as a H-bond donor generally forms one H-bond with the first layer water, as will be shown later in IR spectra. Higher temperatures will promote a faster transformation of these two kinds of structures, increasing the number of surface OH groups between parallel and perpendicular structures. Our results are similar to the previous work reported by Huang et al. with the same functional (PBE) and at the temperature of 400 K, note that the elongated temperature is used for the missing effects of van der Waals effects.[80]

Next, we will discuss the angular distribution of each OH group of every water molecule, the angle is defined as the angle between the OH group and the surface normal. As shown in Fig. 3 (c), the peaks at approximately 70° correspond to the OH groups that are nearly parallel to the surface plane, while the peaks at approximately 170° indicate the OH groups that are nearly perpendicular to the surface. This suggests that more water molecules are oriented towards the interface. DelloStritto et al. confirmed the existence of two types OH groups in the first layer water.[81] These two types OH groups can form H-bonds with two different kinds of surface OH groups mentioned above. The opposite orientation of the second layer water to the first layer water means that more water molecules are oriented up relative to the interface. The orientation of the third layer water is close to disorder, similar to the bulk phase water.

### 3.1.3 Tetrahedral order parameter

To examine the impact of the interface on the local structures of water comparing with the ones of the bulk water, we calculated the orientational tetrahedral order parameter ($Q$) of water. $Q$ characterizes the tetrahedral arrangement of the local structure of liquid water, and it is defined as:[82, 83]

$$Q = 1 - \frac{3}{8}\sum_{j=1}^{3}\sum_{k=j+1}^{4}(\cos\psi_{j,O,k} + \frac{1}{3})^2 \qquad (16)$$

where $\psi_{j,O,k}$ is the angle between the oxygen atom of the center water molecule and two of its four nearest oxygen atoms. A larger $Q$ value indicates a more order water structure, with $Q = 1$ for ice and $Q = 0$ for ideal gas. The results in Fig. 3 (d) illustrates that the structure of water molecules becomes more disordered as they approach the interface. These findings are consistent with the report of Zhang et al.,[54] which shows the $Q$ distribution of interfacial water is generally smaller than that of bulk phase water. From Fig. 3 (e), we can see an overall decrease in the $Q$ distribution values with increasing temperature. Apparently increasing the temperature causes the structure of water to deviate from the tetrahedral. The average value of $Q$ for the bulk phase water at 330 K is around 0.8, which is similar to the results obtained at the PBE-TS level reported in Ref. [81].

### 3.1.4 H-bond network

Finally, we analyzed the number of H-bonds of water molecules along $z$-axis. Here we used the criterium $R$-$\vartheta$ developed in Ref. 35. In this criterium, a H-bond is formed when the intermolecular distance $R$ of two oxygen atoms is smaller than 3.5 Å and the OH…O angle $\vartheta$ is larger than 110°. Fig. 3 (f) displays the average number of total H-bonds ($n_T$), H-bonds donor ($n_D$) and H-bonds acceptor ($n_A$) of each water molecule. In the bulk phase water, $n_T$ is approximately 3.8, and $n_D$ and $n_A$ are the same. As a comparison, in the first layer water, $n_T$ can reach 5, while $n_A$ is slightly larger than $n_D$. This is related to the fact that a partial surface OH groups form H-bonds with the first layer water, with more OH groups acting as donors rather than acceptors when forming H-bonds with the first layer water.

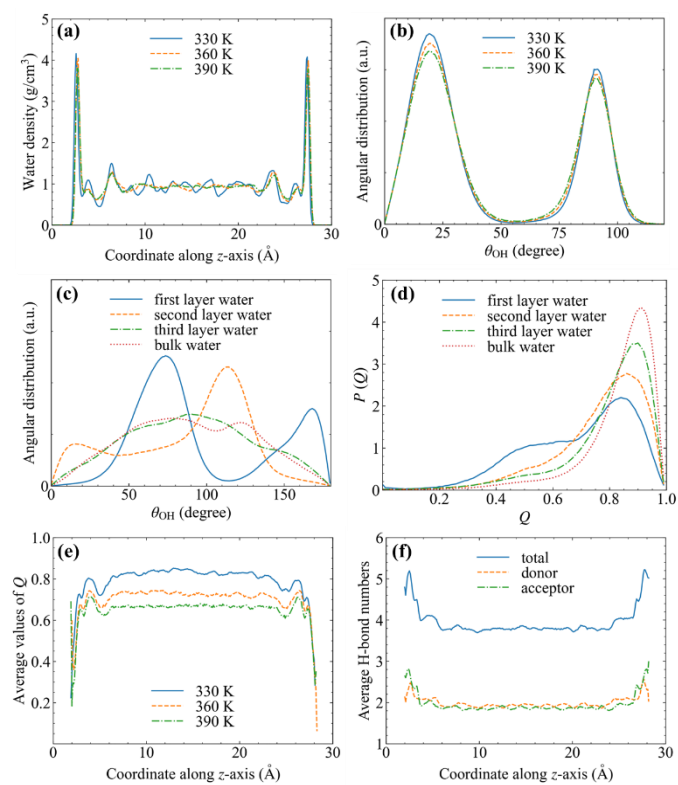

Figure 3. Different statistical values obtained based on MD trajectories for characterizing $\alpha$-$Al_2O_3$(0001)–water interface structure. (a) Density profile of water

along the *z*-axis at different temperatures. (b) Angular distribution of surface OH tilt at different temperatures. (c) Angular distribution of each OH group of each water molecule in different layers makes with the surface normal at 330 K. (d) Average *Q* values distribution of water along the *z*-axis at different temperatures. (e) Distribution of *Q* of water in different layers at 330 K. (f) Distribution of average H-bond numbers of water along the *z*-axis at 330 K.

**3.2 SFG spectra**

The *μ-α* approach was primarily used to calculate SFG spectra in this work. The results were then checked for reasonability using SFG spectra calculated by the ssVVCF approach. In order to facilitate the comparison of the effects of different variables on the SFG spectra, only the imaginary part of the second-order susceptibility with SSP polarization was calculated. In addition, for comparisons that do not involve a change in temperature, the simulation temperature corresponds to 330 K.

**3.2.1    Different cross-correlation cutoff**

Before comparing the results of the *μ-α* approach and the ssVVCF approach with the experimental results, the effects of two factors on the spectra were first determined. First, the cross-correlation cutoff $r_c$ controls the range of intermolecular interactions that contribute to the spectra. When simulating IR and Raman spectra, it is not difficult to converge the spectral shapes by considering the contributions of all intermolecular interactions to the spectra. However, in the simulation of SFG spectra, considering the intermolecular contributions can make the spectral shapes difficult to converge, particularly for the spectra obtained using the *μ-α* approach.[34] Here we plot SFG spectra of the first three layers water with three different values of $r_c$ (0 Å, 3 Å and 6 Å) in Fig. 4 (a) and (b). These values correspond to different conditions: considering autocorrelation only, the interaction between each water molecule and the surrounding first layer water, and the interaction between each water molecule and the surrounding second layer water, respectively.

The impact of the cross-correlation cutoff on SFG spectra obtained using the *μ-α* method is first analyzed. As $r_c$ increases, it can be seen that the two peaks are blue-shifted. Additionally, the intensity of the negative peak at about 3000 cm$^{-1}$ increases, while the intensity of the positive peak at around 3400 cm$^{-1}$ decreases. We specify that the SFG signal for the vibrational mode facing upwards along the interface normal direction is positive. So, the negative peaks correspond to the O-H bond stretching mode towards the interface. Gaigeot et al. believed that the H-bond formed between interfacial water as a donor and surface OH groups is stronger than a normal H-bond.[4] This can explain why it is necessary to consider a cutoff of at least 3 Å in the $\mu$-$\alpha$ approach when the structure contains the H-bonds between interfacial water and surface OH groups. As $r_c$ increases to 3 Å, the spectra change in a similar trend to that given by the $\mu$-$\alpha$ approach, and increases to 6 Å is almost unchanged for the spectra in both methods. This indicates that the intermolecular coupling effects are saturated when the distance beyond 3 Å. Compared with the results obtained by these two methods, we had set $r_c$ = 6 Å to include the intermolecular coupling effects for the rest of analysis.

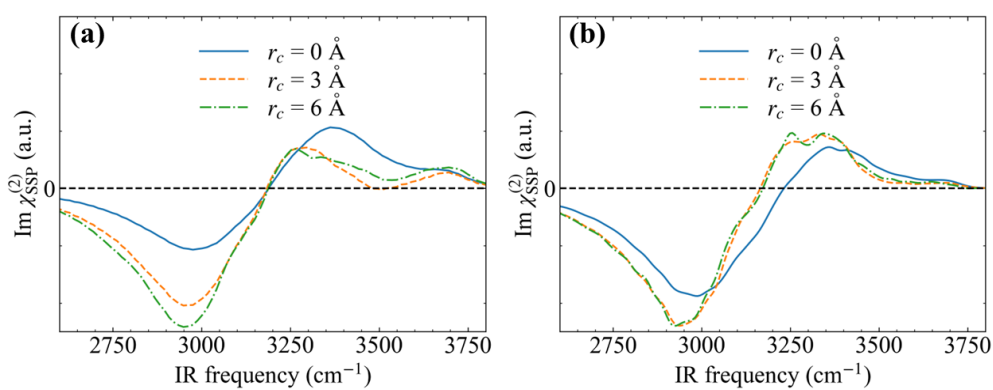

Figure 4. Imaginary part of the theoretical SFG spectra of interfacial water with different $r_c$ using (a) $\mu$-$\alpha$ approach and (b) ssVVCF approach.

### 3.2.2 Impact of the surface on the water molecular response

Now let us focus on the impact of the surface on the water molecular response. Based on the calculated density profile, we decomposed the contributions of different layers water to the SFG spectra, as shown in Fig. 5 (a) and (b). It is clear that both methods give the similar results of decompositions. Obviously the first layer water contributes the most to the SFG spectra, while the second layer contributes significantly less. The third layer water has almost no contribution to the SFG spectra. Compared to Fig. 3 (c), the first- and second-layers water have opposite orientations, but the peaks around 3000 cm$^{-1}$ are all negative. This suggests that the negative peak corresponding to the first layer water mainly comes from the O-H bond with an angular distribution around 170° in Fig. 3 (c), while the negative peak corresponding to the second layer water mainly comes from the O-H bond with an angular distribution around 110° in Fig. 3 (c). All three layers water were taken into account in the later analysis since they all contribute to the SFG spectra.

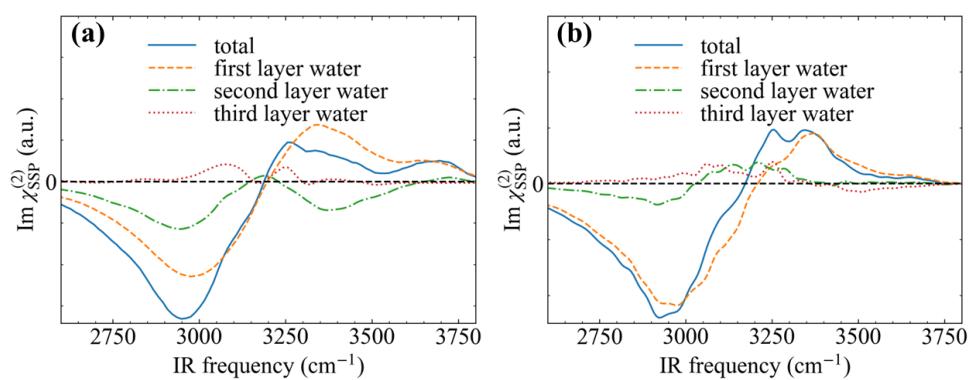

Figure 5. Imaginary part of the theoretical SFG spectra of different layers water using (a) $\mu$-$\alpha$ approach and (b) ssVVCF approach.

### 3.2.3 Temperature effects

The impact of temperature on SFG spectra is now discussed. As shown in Fig. 6 (a) and (b), a higher temperature results in an overall blueshift in the vibrational frequencies of the water. This trend is consistent with the results observed in the previously mentioned IR and Raman spectra. This indicates that even the "up" O-H bond, which corresponds to the positive peak near 3400 cm$^{-1}$, is not a "free" O-H bond, but rather forms a weaker H-bond than the one formed by the "down" O-H bond. In contrast, the vibrational strength remains relatively constant despite changes in temperature. This implies that the orientation of interfacial water is not significantly affected by the change in temperature, as evidenced by the small change in $Q$ of interfacial water with temperature in Fig. 3 (e). Nagata et al. studied the effects of temperature changes on the SFG spectra of the air-water interface.[84] They concluded that the variation of the SFG signal due to temperature change is not related to the orientation of interfacial water. This conclusion is consistent with our findings.

Turning to the surface OH groups, the two methods used to calculate the SFG spectra do not yield identical results, as shown in Fig. 7 (a) and (b). The peak near 3700 cm$^{-1}$ is attributed to perpendicular OH structures that do not form H-bonds. We observed that the corresponding vibrational peaks vary with temperature similarly to the results reported by Nagata et al.[84] The most significant difference between the SFG spectra of surface OH groups predicted by the two methods is the peak at 3400 cm$^{-1}$. In the $\mu$-$\alpha$ approach, this peak has a comparable intensity to the peak at 3700 cm$^{-1}$. However, this peak almost disappears in the results of the ssVVCF approach. It may come from the fact that in the ssVVCF approach, the polarizability $\alpha$ is estimated by using the water molecule while we extend to the surface OH group without considering the effects of Al atoms.

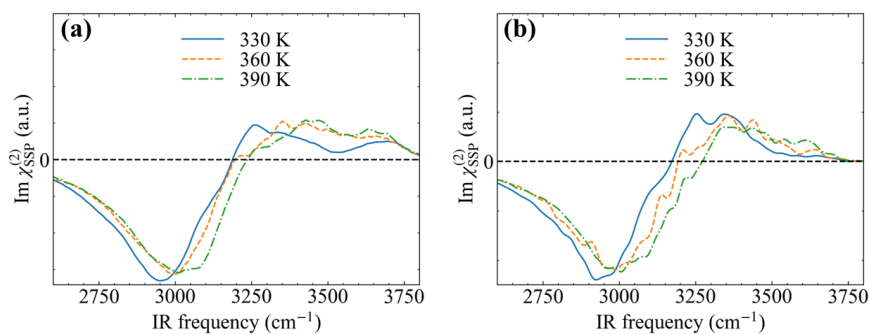

Figure 6. Imaginary part of the theoretical SFG spectra of interfacial water at different temperatures using (a) $\mu$-$\alpha$ approach and (b) ssVVCF approach.

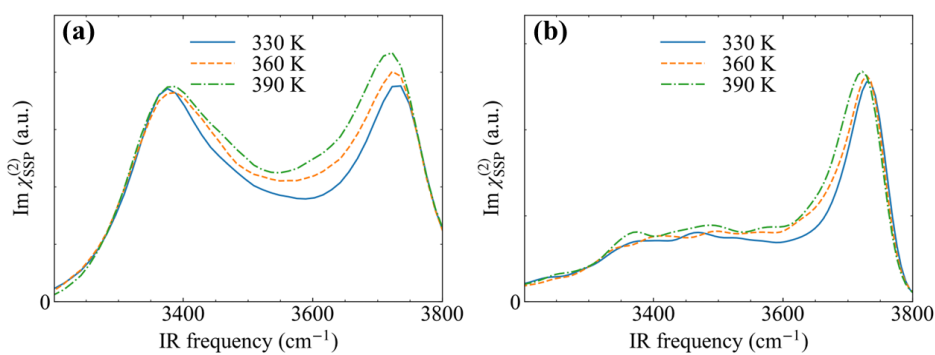

Figure 7. Imaginary part of the theoretical SFG spectra of surface OH groups at different temperatures using (a) $\mu$-$\alpha$ approach and (b) ssVVCF approach.

### 3.2.4 H-bond classification

In the following we investigate the contributions of water forming different H-bond species to the SFG spectra. First, we define a H-bond is formed when the O…O intermolecular distance R is smaller than 3.5 Å and the O-H…O angle is greater than 110°.[35] In the following, we use the labels of "D" and "A", which denote the H-bond donor and H-bond acceptor, respectively. In the calculation, we combined "DA" and "DAA" as one type, as they both contain "free" O-H bonds, which is sufficient for decomposing spectra, For the case of "DAA+DA", we have $n_D \leqslant 1$, which means some small numbers of "A", "D", and "DAAA" species will be taken into account. In the case of "DDA", we have $n_D > 1$ and $n_A \leqslant 1$, this means some small number of "DDDA" will be included. For "DDAA", we have $n_D > 1$ and $n_A > 1$, this means some small number of "DDAAA" and "DDDAA" will be considered. We can see that in Fig. 8 (a) and (b), "DDAA" has a major contribution to the negative peak. Apparently, this is related to the fact that interfacial water has more H-bonds compared with the bulk phase water. "DDA" contributes to the higher frequency region of the negative peak, while the contribution of "DAA+DA" is mainly in the positive peak. Note that "DDAA" also contributes to the positive peak. This could be due to the transformation of "DDAA" to "DAA+DA" within the time window of $t$. Both methods yielded comparable results on the contributions of water molecules with different H-bond types to the SFG spectra.

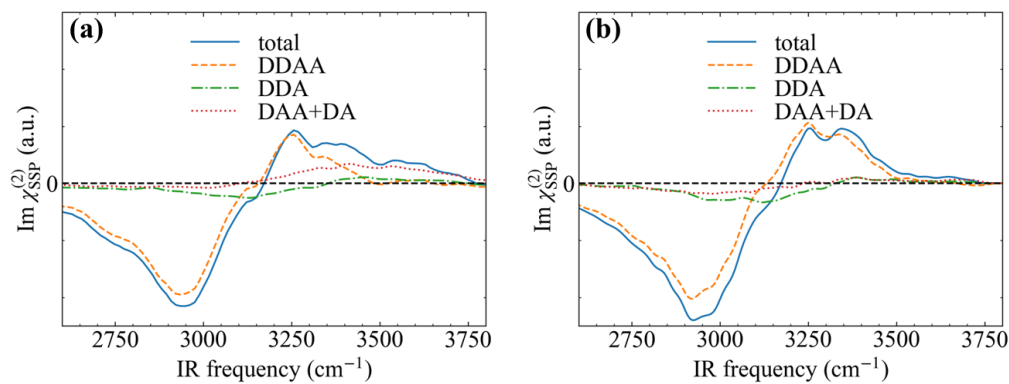

Figure 8. Imaginary part of the theoretical SFG spectra of interfacial water with decomposition of different H-bond types using (a) $\mu$-$\alpha$ approach and (b) ssVVCF approach.

### 3.2.5 Contribution of the surface OH groups

In this section, we would like to discuss the contribution of the surface OH groups to the SFG spectra. As shown in Fig. 7, the spectra of the surface OH groups contain two different peaks. To connect them with the molecular structure, we have decomposed the spectra based on the tilt angle of the surface OH groups, where the tilt angle is defined as the angle between the OH group and the surface normal. The resulting data is presented in Fig. 9. As observed, the surface contains two distinct OH groups: one is a free OH group pointing towards the bulk, while the other forms a H-bond with the bridge oxygen atom of the $Al_2O_3$ surface.

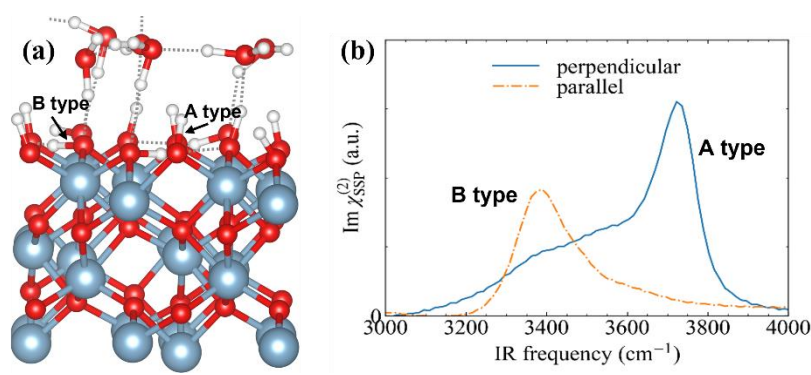

Figure 9. (a) Schematic of different surface OH group at the surface. (b) Imaginary part of the theoretical SFG spectra of different surface OH groups using $\mu$-$\alpha$ approach. Note that surface OH groups with the tilt angle larger than 60° are considered to be perpendicular to the surface (A type), and conversely less than 60° are considered to be parallel to the surface (B type), the tilt angle is defined as the angle between OH group and surface normal.

Next, we would like to comment on the different ratio between 3400 cm$^{-1}$ and 3700 cm$^{-1}$ peaks predicted by $\mu$-$\alpha$ and ssVVCF approaches. The $\mu$-$\alpha$ approach gives a higher 3400 cm$^{-1}$ peak than that predicted by ssVVCF approach, as shown in Fig. 7. To explain it, we calculated the transition dipole and transition polarizability for both surface OH groups and $H_2O$ molecules. The data is shown in Fig. 10. Indeed, the averaged values of $\alpha_{xx}$ and $\alpha_{yy}$ of the surface OH groups are larger than that of the $H_2O$ molecules. Therefore, the difference between ssVVCF and $\mu$-$\alpha$ approaches can be understood as follows: First, the major difference of the spectra comes from the contributions of the surface OH groups with the configurations which are parallel to the surface. In the ssVVCF approach, all the $\alpha_{xx}$ and $\alpha_{yy}$ values of surface OH groups and $H_2O$ molecules are assumed as the same, while, the averaged values of $\alpha_{xx}$ and $\alpha_{yy}$ of the surface OH groups are larger than that of the $H_2O$ molecules. As such, the $\mu$-$\alpha$ approach gives a higher intensity for 3400 cm$^{-1}$ peak than that predicted by ssVVCF approach.

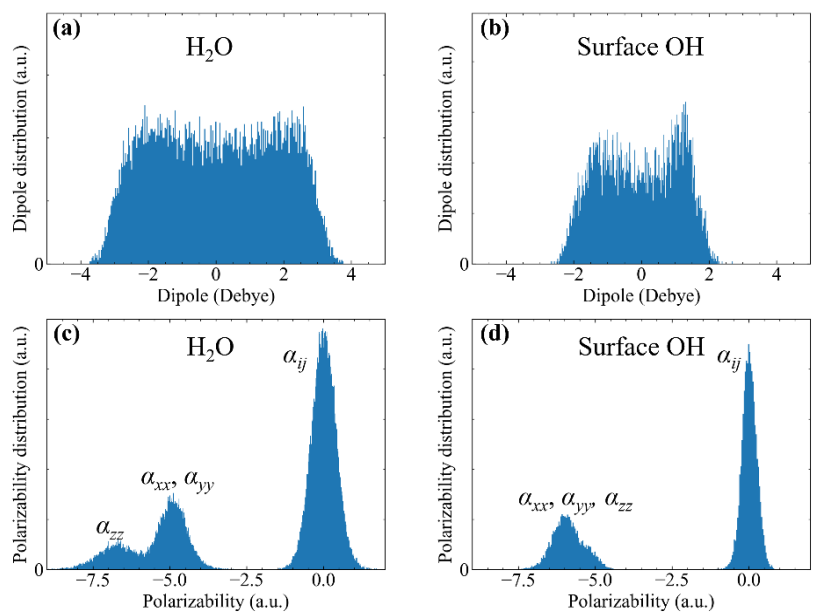

Figure 10. Dipole distribution of (a) water and (b) surface OH groups. The maximum absolute value of the dipole moment of $H_2O$ is twice larger than that of surface OH, because there are two OH groups counted in the calculation of dipole moment calculation. Polarizability $\alpha$ distribution of (c) water and (d) surface OH groups. Note that the overlapping term is $\alpha_{ij}$, where $i, j = x, y, z$ and $i \neq j$. Data of dipole and polarizability comes from the validation data set based on DFT calculations.

### 3.2.6 Comparison with the results in the literature

In this section, we will compare our results with the theoretical and experimental spectra reported in the literature. This will enable a further comparison of the differences between the two methods of calculating SFG spectra.

First, let us focus on the comparison with the theoretical spectra in the literature. Zhang et al. used the ssVVCF approach to simulate the SFG spectra of $\alpha$-$Al_2O_3$ (0001)–water interface.[54] They predicted that the intensity of the positive peak is only one-third of that of the negative peak, which is similar to the conclusions we get using the $\mu$-$\alpha$ approach and ssVVCF approach. Although they use the revPBE functional while choosing $r_c$ = 2 Å, this should not have a significant effect on the intensity ratio of the negative and positive peaks. Melani et al. used the ssVVCF approach to simulate SFG spectra with the same functional (PBE) while setting $r_c$ = 0 Å.[85] The only difference is that their system has a single layer water on the surface. They predicted the same frequencies for the positive and negative peaks, corresponding to 3000 cm$^{-1}$ and 3400 cm$^{-1}$, respectively. Only the signs of the peaks are reversed due to the different reference direction along the interface. Although they predicted essentially the same intensity of positive and negative peaks, their interface model has a significant proportion of "free" O-H bonds. In our model most of the "up" O-H bonds in the interfacial water form H-bonds. The peak intensity at 3400 cm$^{-1}$ will decrease as more water is added to the interface containing a single layer water.

Finally, we discuss the comparison of our theoretical spectra with the experiment. Zhang et al. reported SFG measurements at different pH values of $\alpha$-$Al_2O_3$ (0001)–water interface.[20] They found that for the imaginary part of the second-order susceptibility with SSP polarization, both negative and positive peaks will appear when pH > 7. The shapes of the two methods of the summed spectra corresponding to interfacial water and surface OH groups are closer to the experimental results at the near neutral or slightly basic condition, as shown in Fig. 11 (a) and (b). For the frequencies, the peak of water was significantly red-shifted while the peak of the surface OH groups was essentially the same compared with experiment. This is the known feature of spectra calculated based on PBE functional.[86] As discussed in the previous section, the ssVVCF approach underestimates the transition polarizability components $\alpha_{xx}$ and $\alpha_{yy}$ for the surface groups. Consequently, the 3400 cm$^{-1}$ peak, attributed to surface OH groups with configurations parallel to the surface, is more pronounced in the spectra predicted by the $\mu$-$\alpha$ approach. In our theoretical model, all oxygen sites on the $Al_2O_3$ surface are hydroxylated by H atoms. However, under experimental conditions (pH = 9.0), the reaction $(Al)_2OH + OH^- \rightarrow (Al)_2O^- + H_2O$ deprotonates some surface sites. Therefore, we expect that using fewer hydroxylated

O sites in the theoretical model would reduce the intensity of the 3400 cm⁻¹ peak predicted by the μ-α approach, leading to better agreement with experimental observations. Nevertheless, our theoretical calculations successfully predict the "negative-positive" feature of the SFG spectra, consistent with experimental results near neutral and basic conditions.

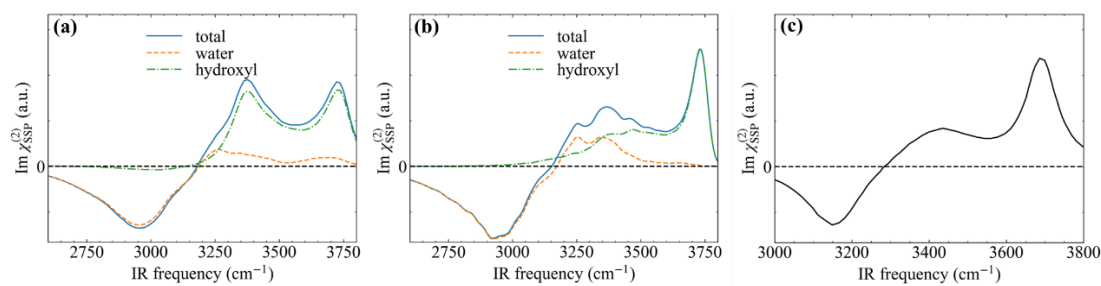

Figure 11. Imaginary part of the theoretical SFG spectra of interfacial water and surface OH groups using (a) $\mu$-$\alpha$ approach and (b) ssVVCF approach. (c) Imaginary part of the experimental SFG spectra of interfacial water and surface OH groups at the basic condition with pH=9.0, the data is reproduced with the permission from Ref. 20. Note that we flipped the sign of the experimental spectra to make the system geometry consistent (i.e. $Al_2O_3$ below water).

## 4. Conclusions

We proposed a framework with the ML models to accelerate MD simulations and vibrational spectroscopy calculations of solid–water interface utilizing $\alpha$-$Al_2O_3$(0001)–water interface as a prototype. Analysis of the trajectory structures and calculation of the vibrational spectra demonstrate that ML models achieve first-principles computational accuracy. We focus on the differences of the SFG spectra based on the $\mu$-$\alpha$ approach and ssVVCF approach. The results indicate that a large cutoff is necessary to calculate the vibrational spectra of $\alpha$-$Al_2O_3$(0001)–water interface when using the $\mu$-$\alpha$ approach. This is due to the significant influence of the interaction between the surface OH groups and the first layer water on the SFG spectra. This effect is distinct from the impact of different cutoffs on SFG spectra at the air-water and hydrophobic interfaces.

Currently, only the SFG spectra of the oxide-water interface under neutral conditions have been calculated in combination with ML models. Calculating the SFG spectra of the oxide-water interface under acidic or alkaline conditions is much more complicated. While the calculation of SFG spectra at different pH at the solid-water interface by AIMD combined with the ssVVCF method has been investigated,[87] we believe that the advantage of ML is that only a small number of configurations need to be added to obtain trajectories corresponding to different pH. Besides, frequent O-H bonds breakage occurs near the interface under acidic or alkaline conditions, which may limit the calculation of SFG spectra based on the ssVVCF method. In contrast, the $\mu$-$\alpha$ method is not affected by the breaking of O-H bonds once the positions of WCs are determined. Nevertheless, it can be expected that ML will accelerate the calculation of SFG spectra for oxide-water interfaces with different pH conditions.

**Acknowledgments**

F. T. is supported by the Natural Science Foundation of Xiamen (grant no. 3502Z202471029) and a startup fund at Xiamen University. J. C. acknowledges the National Natural Science Foundation of China (grant nos. 22225302, 21991151, 21991150, 22021001, 92161113, and 20720220009) and the Laboratory of AI for Electrochemistry (AI4EC) and IKKEM (grant nos. RD2023100101 and RD2022070501) for financial support. This work used the computational resources in the IKKEM intelligent computing center.

**Author Declarations**

Conflict of Interest: The authors have no conflicts to disclose.

**Author Contributions**

**Xianglong Du**: Data curation (lead); Formal analysis (lead); Methodology (equal); Resources (equal); Software (equal); Visualization (lead); Writing – original draft (lead); **Weizhi Shao**: Methodology (equal); **Chenglong Bao**: Methodology (equal); **Linfeng Zhang**: Methodology (equal); **Jun Cheng**: Conceptualization (equal); Project administration (equal); Resources (equal); Writing – review & editing (equal). **Fujie Tang**: Conceptualization (equal); Methodology (equal); Software (equal); Project administration (equal); Writing – review & editing (equal).

**Data Availability**

The data that support the findings of this study are available from the corresponding author upon reasonable request.

**Appendix: IR and Raman spectra of the bulk water**

In this Appendix, we demonstrate the predictive power of our methods, which can accurately reproduce not only interfacial spectra but also the vibrational spectra (IR and Raman) of bulk water. We present the theoretical spectra alongside the corresponding experimental spectra for a direct comparison.

1. **IR spectra**

First, we reported the theoretical IR spectra of the bulk phase water at various temperatures, as shown in Fig. 12 (a). The frequency of the OH stretching mode is blue-shifted, the frequency of the librational mode is red-shifted, and the frequency of the HOH bending mode remains mostly unchanged with increasing temperature, which is consistent with the experimental results.[88]

We then analyzed the contributions of intramolecular and intermolecular interaction to the IR spectra of water to better validate the reliability of our trained ML models, as shown in Fig. 12 (b) and (c). In the bulk phase water, the contributions of intramolecular and intermolecular interactions to IR spectra are comparable. This is similar to the IR spectra decomposition results of Zhang et al.[89] Although the contributions of intramolecular to IR spectra are significantly larger than the intermolecular in the first layer water, it is worth noting that the intramolecular contributions corresponds to a lower vibrational frequency compared with the results in the bulk phase water. This implies that the first layer water forms more H-bonds, which is consistent with the previous analysis of H-bonds.

The surface of $Al_2O_3$ is easy to be hydrated by the OH group, its property may change the absorption behavior of water molecules. Here we calculated the IR spectra of these surface OH groups. The variation of these frequencies with temperature is displayed in Fig. 12 (d). The stretching mode of the surface OH groups corresponds to two peaks larger than 3000 cm$^{-1}$. The peak at the lower frequency is stronger, which is consistent with the findings of Huang et al.,[80] although they replaced hydrogen with deuterium. They concluded that the OH groups that form H-bonds as donor corresponds to the lower frequency peak, while the OH groups that does not form H-bonds as donor corresponds to the higher frequency peak. The former mainly originate from parallel OH structures, while the latter mainly originate from perpendicular OH structures. Additionally, higher temperatures promote the proximity of the two peaks due to the faster transition of the corresponding structures. Therefore, the frequencies of these two peaks become closer as the temperature increases.

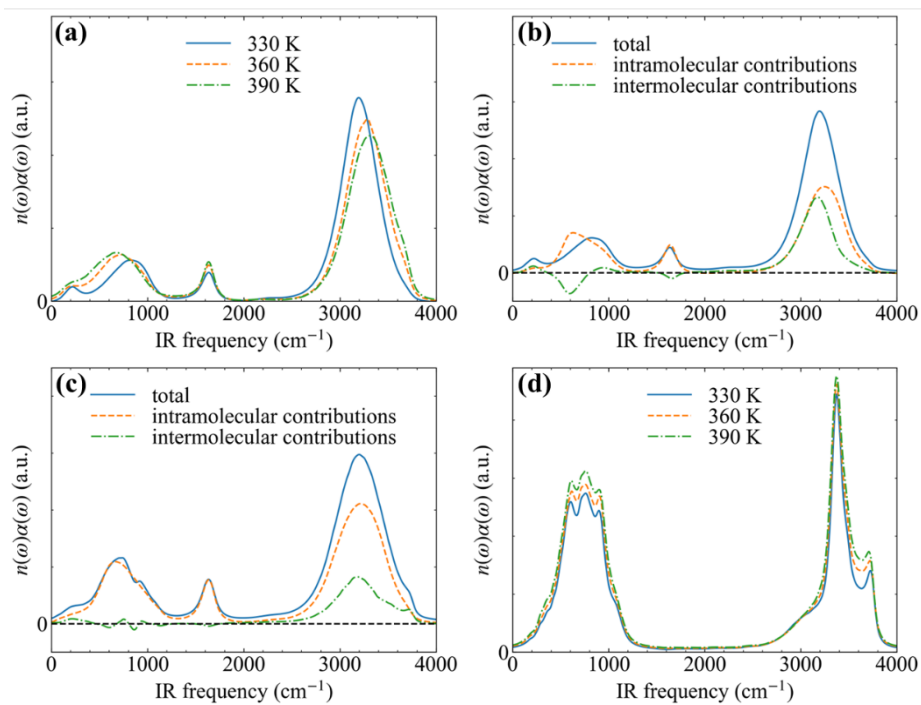

Figure 12. IR spectra of $\alpha$-$Al_2O_3$(0001)–water interface. (a) IR spectra of the bulk phase water at different temperatures. (b) IR spectra of intramolecular and intermolecular contributions of the bulk phase water at 330 K. (c) IR spectra of intramolecular and intermolecular contributions of the first layer water at 330 K. (d) IR spectra of surface OH groups at different temperatures.

### 2. Raman spectra

Here we compared the results with the simulation spectra in the Ref. 45 and experimental spectra in the Ref. [90]. Fig. 13 (a) and (b) show the isotropic and anisotropic Raman spectra of OH stretching mode. In this work, the peaks in the isotropic and anisotropic spectra are at about 3150 cm$^{-1}$ and 3350 cm$^{-1}$ respectively at 330 K. When compared to the frequencies reported in the reference using the SCAN functional at 300 K, they are red-shifted by 100 cm$^{-1}$ and 120 cm$^{-1}$, respectively. That means the frequency difference between the two peaks is basically the same. The difference in predicted frequency is solely due to the different XC functionals used. When comparing the simulation spectra to the experimental spectra in the reference, the effect of temperature variation on frequency is similar.

Next let us move to compare the intramolecular and intermolecular contributions to the isotropic Raman spectra of OH stretching mode. The shapes and frequencies of the predicted peaks are similar to the simulated spectra of the reference when considering the redshift. As the temperature rises, the stretching frequency of the intramolecular contributions is blue-shifted, and the positive and negative peaks corresponding to intermolecular contributions are close to each other. This trend is also consistent with the simulated spectra in the Ref. 45.

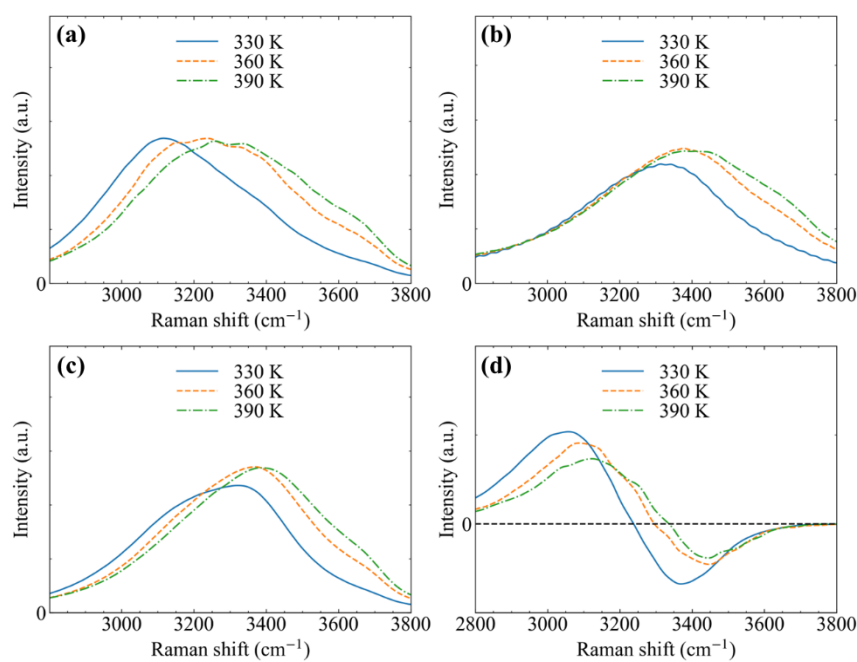

Figure 13. Raman spectra of $\alpha$-$Al_2O_3$(0001)–water interface of the bulk phase water at different temperatures. (a) Isotropic and (b) anisotropic Raman frequency-reduced spectra of water at the OH stretching frequency region. (c) Intramolecular and (d) intermolecular contributions to the isotropic Raman spectra of water at the OH stretching region.

## 3. Comparison with the results in the literature

Finally, we compare our theoretical results with experimental data, as presented in Fig. 14. Overall, the spectral line shapes of the theoretical IR and Raman spectra show good agreement with the experimental data. The observed frequency mismatch is attributed to the used PBE exchange-correlation (XC) functional, which is known to underestimate the OH stretching frequency.[10] Nevertheless, the choice of functional does not affect the temperature dependence of the IR and Raman spectra, as evidenced by the comparison with experimental data in Figs. 12 and 13.

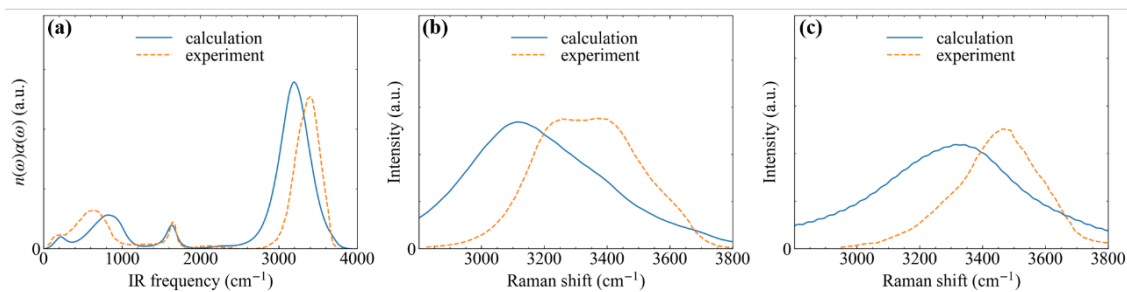

Figure 14. (a) Theoretical and experimental IR spectra of the bulk phase water. (b) Isotropic and (c) anisotropic theoretical and experimental Raman frequency-reduced spectra of the bulk phase water at the OH stretching frequency region. The temperatures of calculation and experiment are 330 K and 300 K. The experimental data of IR and Raman spectra are taken from Ref. 88 and Ref. 90.